\documentclass[twocolumn,showpacs,preprintnumbers,aps,amsmath,amssymb,prb]{revtex4}

\usepackage{graphicx}% Include figure files
\usepackage{dcolumn}% Align table columns on decimal point
\usepackage{bm}% bold math
\usepackage{subscript}
\usepackage{amsmath}
\usepackage{ulem}
\usepackage{color}
\newcommand{\hamilton}{\mathcal{H}}
\newcommand{\matrixX}[1]{\bm{#1}}
\newcommand{\eps}{\varepsilon}
\newcommand{\up}{\uparrow}
\newcommand{\down}{\downarrow}
\newcommand{\tesub}[1]{\textsubscript{#1}}

\newcommand{\op}[1]{\operatorname{#1}}

\begin{document}

\title{Interaction-driven transition between topological states in a Kondo insulator}% Force line breaks with \\

\author{Jan Werner}
\email{jwerner@physik.uni-wuerzburg.de}
\author{Fakher F. Assaad}
\affiliation{Institut f\"ur Theoretische Physik und Astrophysik, Universit\"at W\"urzburg, Am Hubland, D-97074 W\"urzburg, Germany}

\date{\today}

\begin{abstract}
Heavy fermion materials naturally combine strong spin-orbit interactions and electronic correlations.
When there is precisely one conduction electron per impurity spin, the coherent heavy fermion
state is insulating. This Kondo insulating state has recently been argued to belong to the class of
quantum spin Hall states. Motivated by this conjecture and a very recent experimental realization
of this state, we investigate a model for Kondo insulators with spin-orbit coupling. Using DMFT, we
observe an interaction-driven transition between two distinct topological states, indicated by a
\mbox{closing} of the bulk gap and a simultaneous change of the $Z_2$ topological invariant. At large
interaction strength we find a topological heavy fermion state, characterized by strongly renormalized heavy
bulk bands, hosting a pair of zero-energy edge modes. The model allows  a detailed  understanding of the temperature dependence  
of the single particle spectral function and  in particular the energy scales at which one observes the appearance of edge states within the bulk gap.
%Upon reopening of the bulk gap the system
%remains a topological insulator, however with a zero energy edge mode now at the $X$-point
%instead of the $\Gamma-$point.}
\end{abstract}

% PACS, the Physics and Astronomy Classification Scheme.
% candidates:
% 71.10.-w Many-electron systems, theories of; Electronic structure, condensed matter, theories and models of
% 71.70.Ej spin orbit coupling/Stark effect/Zeeman effect/ Jahn-Teller effect in condensed matter
% 71.10.Fd Hubbard model, electronic structure; Lattice fermion models
% 71.27.+a Strongly correlated electron systems; Heavy-fermion solids, electronic states
% 75.30.Mb Fluctuation phenomena, magnetically ordered materials; Heavy-fermion solids, magnetically ordered materials
% 75.70.Tj Magnetic properties, of thin films, surfaces, and interfaces, spin-orbit effects
% 75.20.Hr Mixed-valence solids; Paramagnetism, local moment in compounds and alloys; Heavy-fermion solids diamagnetism and paramagnetism
% 71.28.+d Mixed-valence solids; Narrow-band semiconductors (electron states); Intermediate-valence solids, electron states of
% 71.30.+h Metal-insulator transition; Peierls instability, metal-insulator transitions
% 03.65.Vf Topological phases (quantum mechanics), 

\maketitle
\section{Introduction}
Following the theoretical discovery of the Quantum Spin Hall (QSH) effect
\cite{KM_PRL1, KM_PRL2,Zhang_QSH} and its experimental realization in HgTe quantum well systems
\cite{QSH_exp_Science1, QSH_exp_Science2} the field of topological insulators has attracted a lot of
research interest. Most intriguing is the fact that this state cannot be connected 
adiabatically \footnote{Adiabatically means that no band gap closing or symmetry breaking occurs \cite{book_anderson_cond_mat}.}
to a conventional insulating state. As a consequence, on the boundary between a topological insulator and
a trivial insulator a pair of helical edge modes, which are
robust against disorder and interaction effects, emerges \cite{Zhang_Topo_Ins_RMP,Luttinger_liquid_helical_edge}.
The interplay of correlation effects and topo\-logy has become a very active field of research
\cite{Hohenad_Rev_Topo_Insulators}. Recently a study of the so-called Kane-Mele-Hubbard model
by means of quantum Monte Carlo revealed, that the topological insulator
phase exists up to rather large values of $U$ before undergoing a continuous phase transition to
an antiferromagnetically ordered state, thereby breaking TRS \cite{KMH_PRL, KMH_PRB}.
In the disordered phase the interplay of the collective spin mode
and topology of the band structure leads to novel effects and model systems  \cite{Grover12,Assaad12}.

Hubbard-like correlations have equally been included within dynamical mean-field theory (DMFT) for a variety of models including the
BHZ model \cite{BHZ_Science} of the QSH effect in HgTe quantum 
wells \cite{Corr_Topo_Insulator_BHZ,interaction-induced_TPT_BHZ-model,BHZ_topological_AF,Topological_QPT_QSH}. 
At the DMFT level, collective modes are absent and correlation-driven transitions can be understood in terms of a renormalization of the band structure due 
to a constant real part of the self energy or due to dynamical effects, which can lead to the divergence of the effective mass.
%\comment{Granted, this is a pretty big change. Would you accept it?}
%\xout{
%effects can be understood in terms of a renormaization of the  band structure 
%supplemented by a local,
%Hubbard-like interaction was also studied using DMFT.  With increasing temperature and interaction
%strength the spin Hall conductivity was found to be suppressed \cite{Corr_Topo_Insulator_BHZ},
%eventually leading to a first-order Mott transition. When magnetic ordering was explicitly allow,
%however the Mott transition was masked by a second-order AF transition at smaller U. The
%calculation of the spin Chern number and the AF moment in a subsequent study
%\cite{BHZ_topological_AF} even found a small window where the AF and the TI phase might coexist.
%In a recent investigation, where the effect of including the inter-orbital Hund's exchange J was
%studied systematically \cite{Budich_Topo_Hund_Insulator}, a shift of the transition to larger
%interaction strength was found. Depending on the choice of parameters, a transition from the
%trivial insulator phase at small interaction strength to a topological insulator at larger
%interaction strength, driven by dynamical fluctuations, could be realized. }

DMFT is the method of choice to understand the salient temperature dependent features of the paramagnetic heavy fermion state \cite{Martin10}.
Below the coherence temperature, the individual Kondo screening clouds of the magnetic impurities overlap
coherently to form the heavy fermion liquid. This state of matter is adiabatically connected to the non-interacting system 
and is hence a Fermi liquid. The ultimate signature of coherence in models of heavy fermions is the Kondo insulating state,
where there is precisely one conduction electron paired with an impurity spin \cite{RMP_1D_Kondo_model}.
Prototypical Kondo insulators are YbB\tesub{12}, CeNiSn, Ce\tesub{3}Bi\tesub{4}Pt\tesub{3} and
SmB\tesub{6} \cite{Kondo_insulators_Aeppli}. As proposed in Ref. \onlinecite{Coleman_PRL_TKI} and \onlinecite{Coleman_PRB_TKI} a state with a
non-trivial band topology may be realized in Kondo insulating systems due to the strong spin-orbit
coupling present in these materials, even in the presence of strong electron-electron correlations. In other
words, under certain conditions the Kondo insulating state is adiabatically connected to a
non-interacting insulator with a non-trivial band topology. This results from the odd-parity
wave-function of the f-electrons \cite{Coleman_PRB_TKI}, which leads to a non-trivial,
momentum-dependent hybridization between f-electrons and conduction electrons. Very recently, the
Kondo insulator SmB$_6$ was investigated using an experimental setup suited to identify non-local
transport effects \cite{SmB6_Topo_Exp}. A topological Kondo insulating state was indeed found to exist
\cite{SmB6_Topo_Exp, SmB6_Spectroscopy_Exp, SmB6_Surface_Hall_Exp}, explaining very naturally both
the residual conductivity as $T \to 0$ and the existence of in-gap states found in ARPES
measurements \cite{SmB6_in_gap_states}. Follow-up investigations using a similar method may very
well find topologically non-trivial states in other Kondo insulating materials, which are up to now
not very well understood \cite{SmB6_Topo_Exp}.%

The aim of this paper is to investigate models of heavy fermions within the DMFT framework which have 
a topological Kondo insulating ground state. One of our central interests is to assess if this state of matter can 
occur in the strong coupling local moment regime. In section \ref{sec_TKI_model} we describe our model for topological Kondo insulators. Next we
briefly introduce the concept of the topological invariant and explain its evaluation for an
interacting system, using the so-called topological Hamiltonian. After describing our method in
section \ref{sec_Method}, we determine the topological properties of the model. In sections
\ref{sec_Transition} and \ref{sec_heavy_band} we
present our main findings, before concluding  in section \ref{sec_Discussion}.

\section{Topological Kondo Insulator}
\label{sec_TKI_model}
We will devise a model appropriate for Ce-based compounds in 2D \cite{Reinert_HF_CePt5}. Our starting point is the Ce ion's J=5/2
multiplet, which originates from the 4f-orbitals in the presence of spin-orbit coupling. In the
solid state, the multiplet is split further by the crystal electric field, with the splitting
depending on the lattice symmetry\cite{Coleman_PRB_TKI}. We assume that the ground state of the
magnetic ion is a Kramer's doublet, which we call f-states. In the simplest case, they hybridize with a
single conduction band \mbox{(d-states)}. With these ingredients, we build a model quite
similar to the Periodic Anderson Model (PAM) on a 2D square lattice. The resulting Hamiltonian is
\begin{align}
  \label{eq_hamiltonian}
  \hamilton &= \hamilton_0 + \hamilton_U\\
  \hamilton_0 &= \sum \limits_{k\in BZ}
  \begin{pmatrix} d^{\dagger}_{k} \\ f^{\dagger}_{k} \end{pmatrix}^{T}
  \begin{pmatrix}
     \matrixX{E_d}(k) &  V^{\ast} \matrixX{\Phi}^{\dagger}(k) \\
    V \matrixX{\Phi}(k) & \matrixX{E}_f(k)
  \end{pmatrix}
  \begin{pmatrix} d_{k} \\  f_{k} \end{pmatrix} \notag\\
  \hamilton_U &= U \sum \limits_i n^{(f)}_{i\up} n^{(f)}_{i\down} \notag
\end{align}
where
$\matrixX{\Phi}(k), \matrixX{E}_d(k)$ and $\matrixX{E}_f(k)$ are 2x2 sub-matrices, and encode
all the information on the geometry and the effect of spin-orbit coupling. We use
$( d^{\dagger}_{k} \, f^{\dagger}_{k} )$ as a short hand notation for
$( d^{\dagger}_{k \up} \, d^{\dagger}_{k\down} \, f^{\dagger}_{k \up} \, f^{\dagger}_{k \down})$,
creating an electron in the conduction band and the almost localized band, respectively. Here $\up$
and $\down$ denote the pseudo-spin quantum number.

Due to spin-orbit coupling the f-states are
eigenstates of the total angular momentum J, and hence hybridize with conduction band states with
the same symmetry. This gives rise to the momentum-dependence and non-trivial orbital structure
of the form-factor $\matrixX{\Phi}(k)$ \cite{Coleman_PRB_TKI}. Following the derivation in Ref.
\onlinecite{Tran_PRB_TKI}, we obtain 
\begin{align*}
\matrixX{E}_d(k) &= -2 t (\cos(k_x) + \cos(k_y)) \matrixX{1}\\
\matrixX{E}_f(k) &= \Bigl ( \eps_f-2 t_f (\cos(k_x) + \cos(k_y)) \Bigr ) \matrixX{1}\\
\matrixX{\Phi}(k) &= \vec{d}(k) \circ \vec{\matrixX{\sigma}}\\
\text{where } \quad d(k) &= ( 2 \sin(k_x), 2 \sin(k_y),0 )
\end{align*}
formulated in the basis of the Pauli matrices $\vec{\matrixX{\sigma}}$ and the 2x2 unit matrix
$\matrixX{1}$.  The fact that $\matrixX{E}_{d,f}(k) $ ($\vec{d}(k)$) is even (odd) under the transformation $k \rightarrow -k $ guarantees 
time reversal symmetry. 
The interaction term $\hamilton_U$ acts only on the almost localized f-states. The
delocalized conduction electrons are only weakly affected by correlation effects, therefore we
neglect interactions on the d-orbitals. We let the hybridization amplitude take a moderate value of
$V = 0.4 \,t$, and allow for a small, hole-like hopping amplitude between f-orbitals, i.e.
$t_f = -0.2 \, t$.
% \comment{I actually do not know what a hole like hopping means}
Both will be renormalized to a smaller value in the presence of the
interaction.\\
The non-interacting part of our model can be related to the BHZ model
\cite{BHZ_Science,interaction-induced_TPT_BHZ-model} for HgTe quantum wells exhibiting the QSH effect.
It is given by
\begin{align}
\label{eq_hamiltonian_BHZ}
\hamilton(k) = & \begin{pmatrix} \hamilton_{BHZ}(k) & 0 \\ 0 & \hamilton^{\ast}_{BHZ}(-k) \end{pmatrix} \\
\hamilton_{BHZ}(k) = & \Bigl ( m - \cos(k_x) - \cos(k_y) \Bigr ) \matrixX{\sigma}_z \notag \\
&+ \lambda \Bigl ( \sin(k_x) \matrixX{\sigma}_x + \sin(k_y) \matrixX{\sigma}_y \Bigr)\notag 
\end{align}
which can be obtained from \eqref{eq_hamiltonian} in the mixed-valence limit $t_f = -\, t$. The
parameters are related by $\lambda = V$ and $m = -\eps_f/4$. Thus we expect some similarities especially in the weak coupling 
limit. Our primary interest however is the correlation-dominated Kondo insulating state.

\section{Topological Invariant}
\label{sec_Topo_Inv}
The topological insulator does not connect adiabatically to the trivial band insulator \cite{Zhang_Topo_Ins_RMP}.
Both states cannot be characterized by a local order parameter and one has to resort to  the concept of a
global $Z_2$ topological invariant $\nu=0, 1$ to distinguish them 
\cite{KM_PRL1}. Provided that the protecting time reversal symmetry is not broken, states
characterized by different values of $\nu$ can't be adia\-ba\-ti\-cally connected without closing the
single particle gap. A formulation of the topological invariant in terms of the single-particle
Green function \cite{Zhang_Topo_Inv,Zhang_Topo_Simp_Inv} quite naturally extends the concept to
interacting systems adiabatically connected to the
non-interacting case\cite{Budich_Topo_Stat_Mat}. It turns out that the zero-frequency value of the
Green function suffices to characterize the state in terms of the topological invariant
\cite{Zhang_Zero_Freq_Green}, which leads to the idea of defining a so-called topological Hamiltonian
\cite{Topo_Hamiltonian}
\begin{equation}
\label{eq_h_topo}
h_{topo}(k) = - G^{-1}(k, i \omega = 0) = h_0(k)+\Sigma(k, i \omega = 0).
\end{equation}
The topological invariant of the interacting system is given by the topological invariant
corresponding to the non-interacting system governed by the topological Hamiltonian $h_{topo}$.
It characterizes correctly the system in terms of the topological invariant, but it doesn't provide
realistic quasi-particle spectra \cite{Topo_Hamiltonian}. In the presence of inversion symmetry, the
calculation of the topological invariant of $\eqref{eq_h_topo}$ is even further simplified
\cite{Topo_Ins_Inversion_Symm}. One simply has to calculate the quantity
\begin{equation}
\delta_i = \prod_m \xi_{2m}(\Gamma_i)
\end{equation}
where m is the band index, $\Gamma_i$ are the four time reversal invariant momenta in 2D,
namely $\Gamma$, $M$ and the equivalent $X$ and $Y$. $\xi_{2m}$ is the corresponding
parity eigenvalue at $\Gamma_i$ of the 2m-th band. The $Z_2$
topological invariant is then obtained via
\begin{equation}
\label{eq_Z2_inversion}
(-1)^{\nu} = \prod_i \delta_i.
\end{equation}
Besides topological and trivial insulators, which differ in the $Z_2$ index $\nu$, in a lattice system, due to the presence
of the space-group symmetry, distinct phases can appear, which share the same value of $\nu$ \cite{TI_space_group}.
However, they differ in the values of the individual parities $\delta_i$ at the time-reversal invariant momenta $\Gamma_i = (\Gamma,M,X,Y)$.
As these parities are conserved, the corresponding phases are separated by a quantum phase transition
where the bulk band gap closes. In particular, the BHZ model on the square lattice
exhibits two distinct, topologically non-trivial phases. The $\Gamma$-phase with the
parities $\delta_i = (-1,1,1,1)$ is characterized by a skyrmion structure of the Berry phase centered at zero momentum,
while the $M$-phase with $\delta_i=(1,-1,1,1)$ exhibits a Berry phase skyrmion at
finite momentum, and was therefore termed a "translationally active" topological insulator in Ref. \onlinecite{dislocation_pi_flux_TI}.
It is only in this phase, that dislocations can act as a $\pi$-flux, binding zero-energy modes.
In addition, the model may exhibit an insulating phase associated with the equivalent X and Y points
with a trivial band topology, which is in fact realized by introducing a long-range hopping \cite{TI_space_group}.

\section{Method}
\label{sec_Method}
\begin{figure}[b]
	\includegraphics[width=0.483\textwidth]{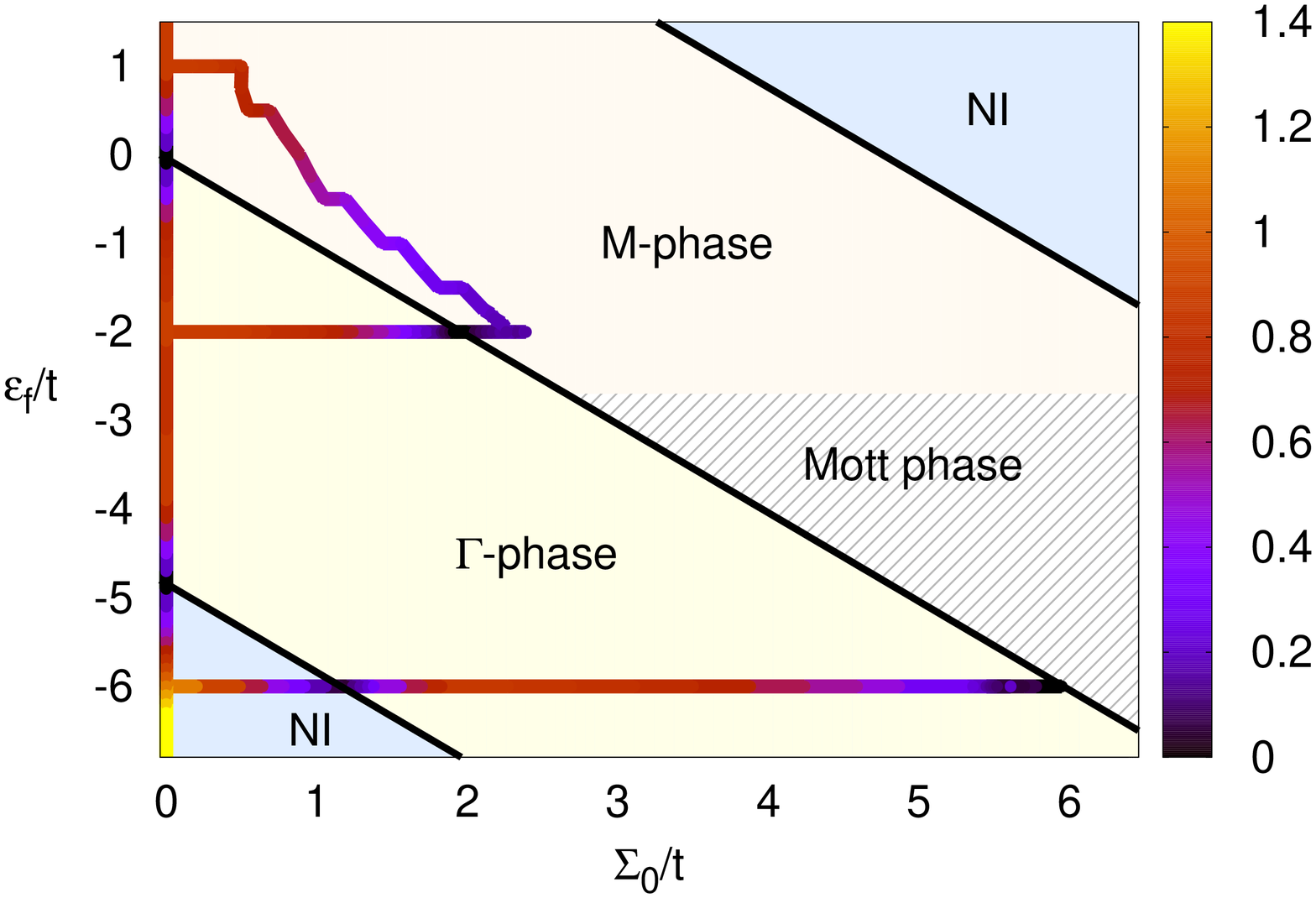}
	\caption{Phase diagram of the topological Hamiltonian \eqref{eq_h_topo_simp} for
	$t_f = -0.2  \,t$. On the lines where the right-hand side of equation \eqref{chern_explicit}
	evaluates to zero, the topological invariant is not defined, and the system is in a gap-less,
	semi-metallic state. The colored, thick lines represent the simulation runs for different
	$\eps_f$ and $U$ and the analytical results at $U=0$, were $\Sigma_0 = 0$. The line color
	corresponds to the band gap size $\Delta_g$. The actual gap closings coincide very well with
	the predicted transition lines. In a certain part of phase space, instead of the $M$-phase,
	a possible non-topological Mott phase was found. There the colored lines end. NI denotes the trivial insulator.}
	\label{fig_chern}
\end{figure}

We employ the DMFT to make our model system \eqref{eq_hamiltonian}
accessible to numerical methods \cite{RevDMFT96}. The central approximation of the DMFT is the
assumption of a momentum-independent self energy
\begin{equation}
\label{approx_dmft}
\Sigma(k, i \omega) \sim \Sigma(i \omega).
\end{equation}
This approach neglects non-local correlations. In addition, as we are interested in the properties
of the para\-mag\-netic phase, we do not allow for magnetically ordered states, which 
would break TRS.
%\xout{As we are interested in the properties of the paramagnetic phase, and
%given the purely local interaction, the local approximation is well justified.}
%\comment{ The fact that the  Hubbard U is local does not mean that DMFT is justified.
%A Hubbard U can generate non-local physics which cannot be captured at the DMFT level!}
This translates to a simplification of the topological Hamiltonian \eqref{eq_h_topo}, which is
now the original Hamiltonian renormalized by a constant
\begin{equation}
\label{eq_h_topo_simp}
h_{topo}(k) = h_0(k)+ \Sigma(i \omega = 0) = h_0(k) + \Sigma_0.
\end{equation}
We can now find an explicit formula for the $Z_2$ topological invariant \eqref{eq_Z2_inversion}
in terms of our model's bare parameters and the renormalization constant $\Sigma_0$:
\begin{equation}
\label{chern_explicit}
\begin{split}
(-1)^{\nu} =& \op{sign}(\eps_f+\Sigma_0)^2 \times \op{sign}(4 t-4 t_f-\Sigma_0-\eps_f) \times \\
&\times \op{sign}(4t_f-4 t-\Sigma_0-\eps_f)
\end{split}
\end{equation}
The resulting phase diagram in terms of $\eps_f$ and $\Sigma_0$ is shown in Fig.
\ref{fig_chern}. It represents a roadmap, which simplifies the characterization of the system's
state once $\Sigma_0 = \Sigma_0(U)$ is known.
The simplification \eqref{approx_dmft} allows for a mapping of the lattice problem
\eqref{eq_hamiltonian} to an auxiliary single-impurity problem, which describes a fully correlated
impurity $\hamilton_{imp}$ coupled to a non-interacting bath. It can be formulated in terms of an
effective action
\begin{equation}
\label{eq_S_eff}
S_{eff} = \int d \tau d \tau' f^{\dagger}(\tau) \matrixX{\Delta}(\tau-\tau') f(\tau') + \int d \tau\, \hamilton_{imp}(\tau).
\end{equation}
The bath parameters $\matrixX{\Delta}(\tau)$ have to be obtained in a self-consistent manner. Once
convergence is reached, the impurity self-energy is equivalent to the local self-energy
$\Sigma(i \omega)$ of the original system.\\
We use the numerically exact CT-HYB quantum Monte Carlo algorithm \cite{PWernerPRL,PWernerPRB}
as the impurity solver at low but finite temperature, which yields the self-energy in \mbox{imaginary}
frequency. The inverse temperature was chosen as $\beta t = 100$ in most cases, while close to the
transitions it was increased up to $\beta t = 300$.\\
In the Fermi liquid regime, $\op{Im}[\Sigma(i \omega)] \sim i \omega$, while the real part
can be extrapolated reliably to a finite value, i.e. $\op{Re}[\Sigma(i \omega \to 0)] = \Sigma_0$.
From the local one-particle Green's function in imaginary time $G_{loc}(\tau) \simeq \exp(\tau \eps_{\pm})$
we infer the position of the upper and lower band edges $\eps_{\pm}$. This method fails when the band gap size
becomes comparable to or smaller than the finite simulation temperature.
%The extraction of the band gap size becomes unreliable for $\Delta_g \lessapprox 0.05\, t$.

\section{Transition between distinct topological states}
\label{sec_Transition}
\begin{figure}[htbp]
	\includegraphics[width=0.483\textwidth]{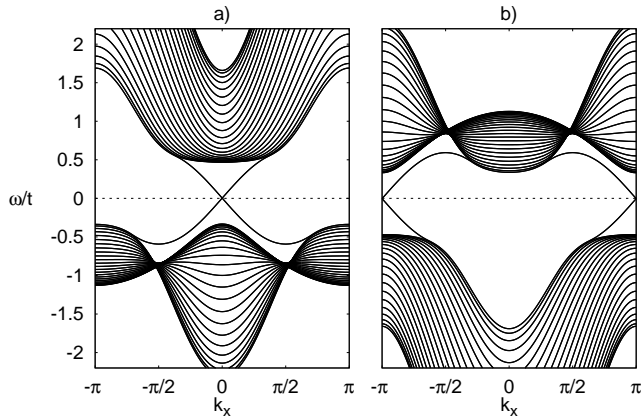}
	\caption{Band dispersion for the non-interacting case with open boundaries in
	y-direction, $N_y = 24$, a) crossing at $\Gamma$, b) crossing at $X$.}
	\label{fig_nonint_edge_modes}
\end{figure}

\begin{figure}[htbp]
	\includegraphics[width=0.483\textwidth]{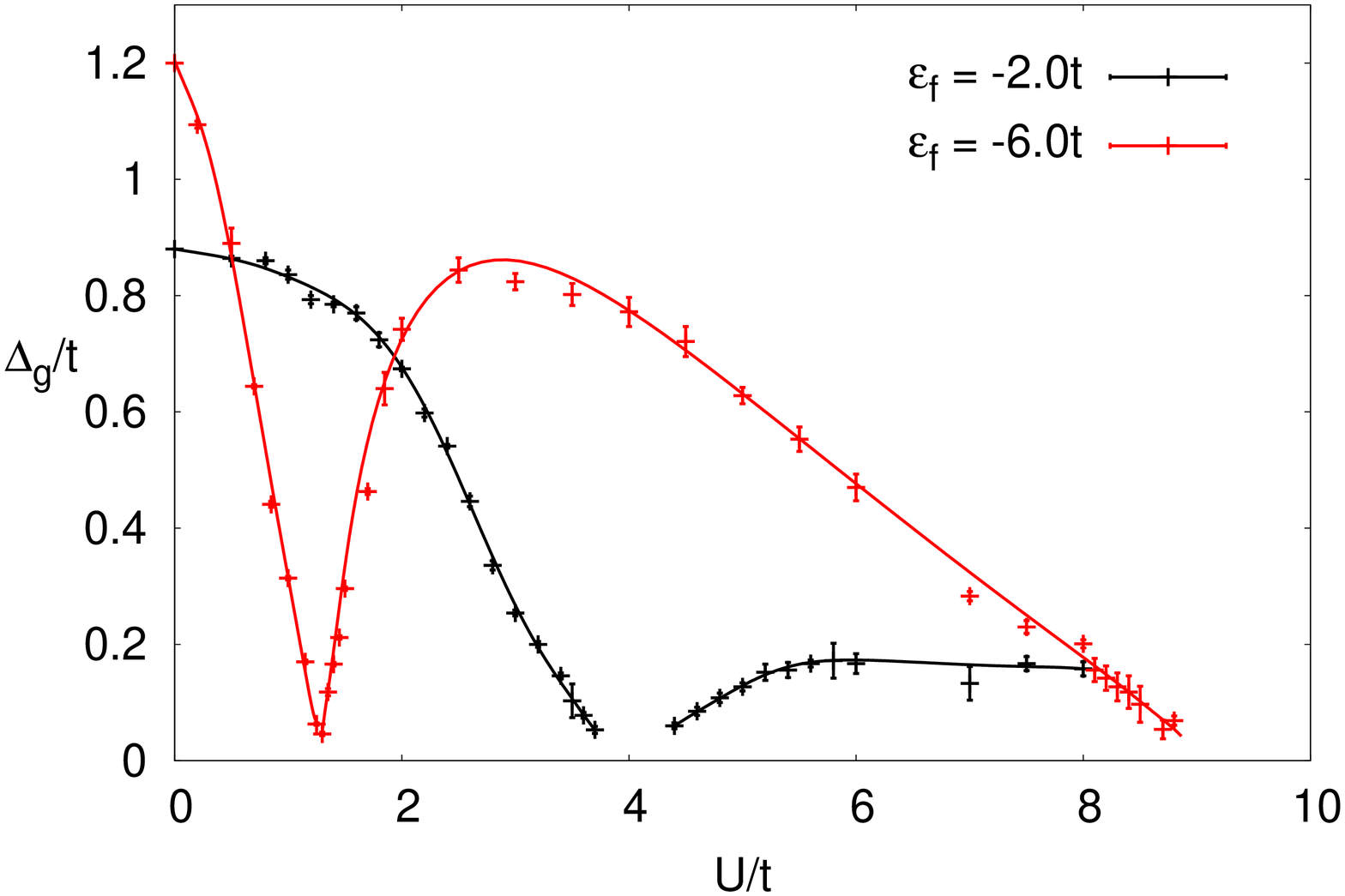}
	\caption{Evolution of the bulk band gap with increasing U for different values of
	$\eps_f$. The solid curves are guides to the eye.}
	\label{fig_ef2_band_gap}
\end{figure}

In Fig. \ref{fig_chern} the phase diagram of the topological Hamiltonian is shown, superimposed
with numerical results for the evolution of the band gap size $\Delta_g$ as a function of
$\Sigma_0(U)$, for different values of $\eps_f$. Positioning the bare f-level at
$\eps_f = -2.0 \,t$ puts the system in a topological insulating state
at $U = 0$, with a bulk band gap $\Delta_g \approx 0.88 \,t$. On a geometry with open boundaries
or with an interface to a conventional insulator, this leads to the formation of a pair of
zero-energy edge modes inside the bulk gap. They cross at the $\Gamma$-point $k_x = 0$, as can be
seen in Fig. \ref{fig_nonint_edge_modes}a. This state is equivalent to the $\Gamma$-phase of the BHZ-model.
As we slowly increase $U$, the system stays in this phase
as long as the bulk gap remains open and TRS is not broken spontaneously.
The results for the bulk band gap size $\Delta_g$ are shown in Fig. \ref{fig_ef2_band_gap}. While
increasing $U$, we adapt the position of the chemical potential $\mu$ to always
stay approximately in the center of the band gap. This way we also ensure, that the system is
precisely half-filled.
Approaching $U=4.0\,t$, the size of the band gap smoothly decreases, and the gap eventually closes.
For this parameters, the band gap is too small to be reliably calculated with the above method. As a result,
the values for the band gap size around $U = 4.0\,t$ were not included in Fig. \ref{fig_ef2_band_gap}.\\
Right at $U = 4.0 \,t$ the system is in a semi-metallic state where the topological invariant is not
defined, and can therefore change discontinuously while
crossing this point in phase space. In the phase diagram of Fig. \ref{fig_chern}
this point is exactly on the black, singular line given by $\Sigma_0 = -\eps_f$. Increasing
U even further leads to a reopening of the band gap.
The calculation of
the topological invariant using the topological Hamiltonian \eqref{eq_h_topo_simp} reveals, that
for $U > 4.0 \,t$ the system is still in a topological insulating state. Not surprisingly, it is
distinct from the state at $U = 0$. Due to the band gap closing at $U = 4.0 \,t$, an adiabatic
connection to the $\Gamma$-phase is not possible. Instead this state is connected to a different
topological state, namely the one present for $\eps_f > 0$. This state is in fact equivalent to the $M$-phase of the
BHZ-model. In this case the edge states cross at $k_x = \pm \pi$, as shown in Fig.
\ref{fig_nonint_edge_modes}b. The adiabatic connection can indeed be established by smoothly
changing $\eps_f$ and $U$ in a way, such that the system stays in this phase. This is done in
a simulation run starting at $\eps_f = -2.0 \,t, \, U = 6.0 \,t$, which corresponds to
$\Sigma_0 \approx 2.3 \,t$, and evolving the parameters to $\eps_f=+1.0 \, t$ and $U=0$, while staying well
away from the singular line $\Sigma_0 = -\eps_f$. Along this path we find no gap closing, as
the upper-most colored, somewhat wiggly line in Fig. \ref{fig_chern} clearly shows.
We explicitly calculated the parities $\delta_i$ and found agreement with the combination
of parities of the $\Gamma$-phase and the $M$-phase, respectively.
This signifies that at $U = 4.0 \,t$ we observe a transition between these two distinct topological
states, which is driven by the interaction.\\

\section{Heavy band topological insulator}
\label{sec_heavy_band}
\begin{figure}[htbp]
	\includegraphics[width=0.483\textwidth]{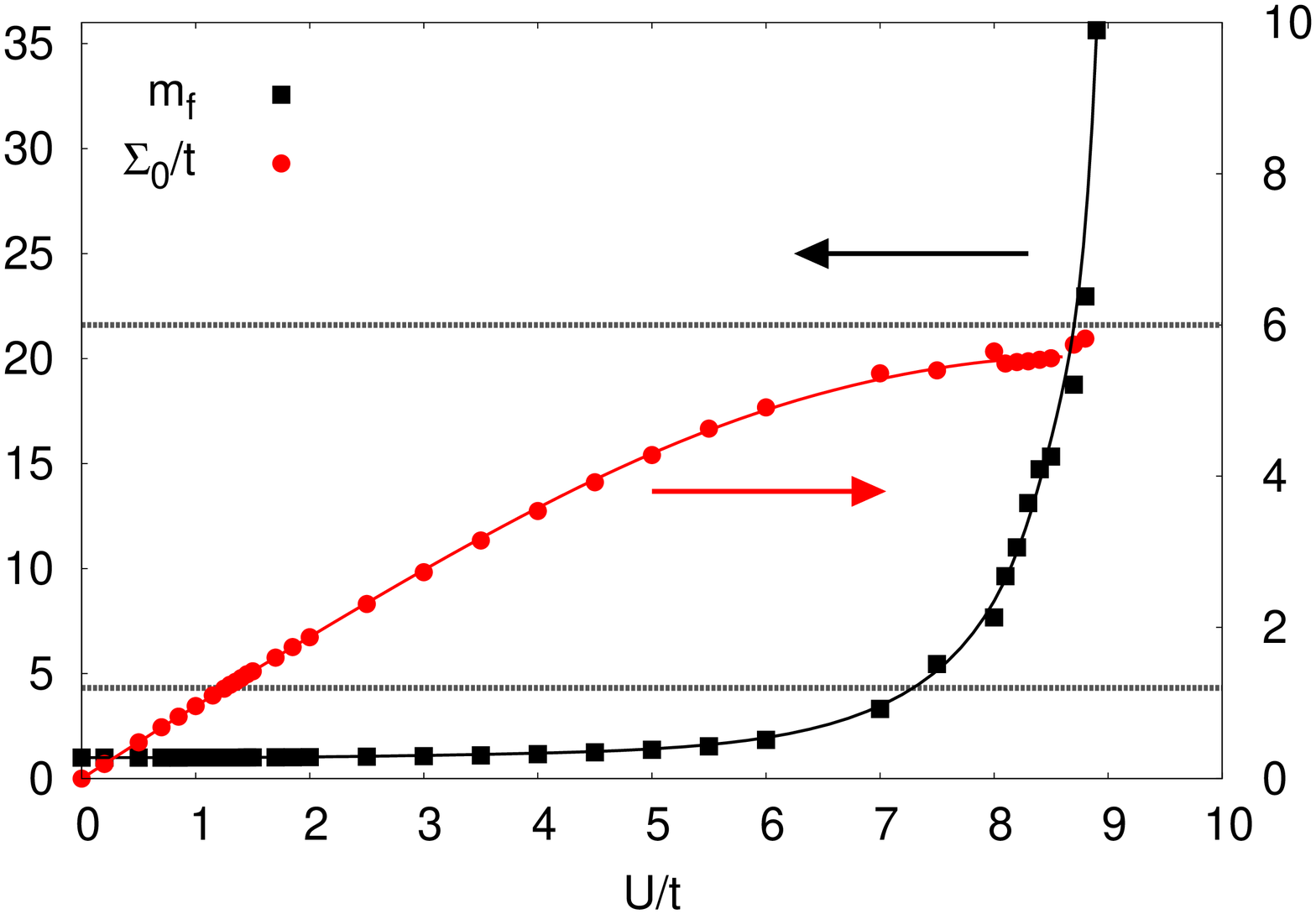}
	\caption{Evolution of the renormalization constant $\Sigma_0$ and the effective mass
	$m_f$ with increasing U for $\eps_f = -6.0 \,t$. The horizontal lines indicate the
	predicted transition lines to the $\Gamma$-phase at $\Sigma_0 = 1.2 \,t$ and the $M$-phase at
	$\Sigma_0=6.0\,t$ obtained from \eqref{chern_explicit}. The solid curves are guides to the
	eye.}
	\label{fig_ef6_m_eff}
\end{figure}

One can think of other paths in the phase diagram of the topological Hamiltonian. For the choice
$\eps_f = -6.0 \,t$ the non-interacting system is an insulator with a \mbox{trivial} band topology. The weakly
dispersing f-band is se\-pa\-ra\-ted from the d-band by a band gap $\Delta_g \approx 1.2 \,t$. Smoothly
increasing $U$ leads to a phase transition to the $\Gamma$-phase once the transition line in Fig.
\ref{fig_chern}, defined by $\Sigma_0 = 4t_f-4t-\eps_f$ is reached, which is the case at
$U \approx 1.25\, t$. Indeed, this transition is the one observed in Ref.
\onlinecite{interaction-induced_TPT_BHZ-model} and \onlinecite{Budich_Topo_Hund_Insulator}
in the BHZ-Hubbard model for $m>2$ at weak coupling
strengths. Beyond the semi-metallic transition point, as the system is driven deeper into the
topological insulator phase, the band gap size increases. It starts to decrease again beyond
$U=3.0 \,t$. For values of $U > 8.0 \,t$ the bulk band gap size becomes very small, signaling the
proximity to the next transition line, which separates the two distinct topological insulator
states of $\Gamma$-phase and $M$-phase. Fig. \ref{fig_ef6_m_eff} reveals that the renormalization constant
$\Sigma_0(U)$ depends on $U$ in a non-linear way. As we increase $U$ beyond
the value of $U = 8.0 \,t$, the evolution of $\Sigma_0$ becomes
very flat. While it approaches the value of $\Sigma_0 = 6.0 \,t$, where the transition should
happen, the effective mass, as defined by
\begin{equation}
\label{eq_m_eff}
m_f = 1-\frac{d \op{Im} \Sigma(\omega)}{d \omega} \Bigr \vert_{\omega \to 0} \approx 1-
\frac{\op{Im} \Sigma(i \omega_0)}{\omega_0}
\end{equation}
increases rapidly, and apparently diverges at $U \approx 8.9\,t$. As a consequence,
$\op{Im}[\Sigma(i \omega)]$ remains finite in the limit $\omega \to 0$, indicating a possible
orbital-selective Mott transition of the heavy band due to local dynamical fluctuations \cite{FDWN_interactions_TI},
or a transition to an antiferromagnetic state,
which can't be captured in our approach. The possibility of a TRS breaking transition is supported by
the fact that the f-band is exactly half-filled at this point, while the double occupancy is strongly suppressed. The nature of the
transition and the phase lying beyond the transition can't be investigate with our current method.
Most probably, due to the approximation of DMFT, in our simulations the size of the band gap
becomes so small that it can't be resolved, but it never closes \cite{3HeModel2011,interaction-induced_TPT_BHZ-model}.

\begin{figure}[htbp]
	\includegraphics[width=0.483\textwidth]{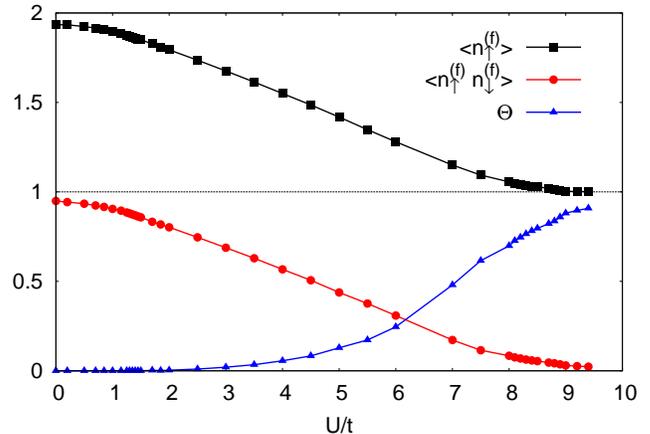}
	\caption{f-level occupancy, double occupancy and the measure of correlation $\Theta$ for
	$\eps_f=-6.0\, t$ as a function of U.}
	\label{fig_ef6_corr}
\end{figure}

\begin{figure}[htbp]
	\includegraphics[width=0.483\textwidth]{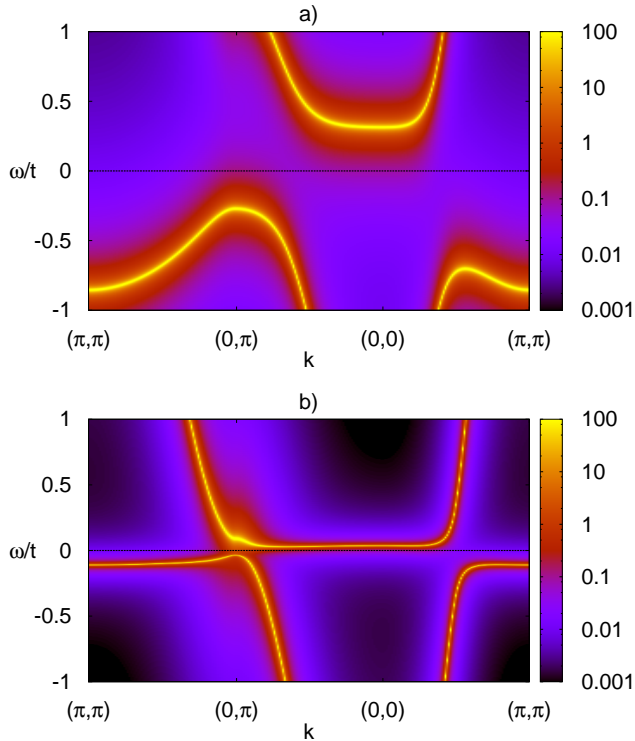}
	\caption{Low energy part of the spectral function for $\eps_f = -6.0\, t$, with
	an artificial broadening $\delta = 0.01$. The interaction strength is a) $U = 5.0\,t$, with a resulting band gap
	$\Delta_g \approx 0.63\,t$ and b) $U = 8.2\,t$ with $\Delta_g \approx 0.14\, t$. Clearly
	visible is the formation of the heavy bands in the strongly correlated case.}
	\label{fig_Aom}
\end{figure}

To quantify the importance of
correlations, and especially to pin down the local moment regime, it is interesting to investigate the quantity
\begin{equation}
\label{eq_corr}
\Theta = 1-\frac{\langle n^{(f)}_{\up} n^{(f)}_{\down} \rangle}{\langle n^{(f)}_{\up}
\rangle \langle n^{(f)}_{\down} \rangle}
\end{equation}
where $\Theta = 0$ without correlations. For a local moment, where the double occupancy is
completely suppressed, $\Theta = 1$. Thus we can use this quantity as a measure for the impact of
correlations on the system. The evolution of $\Theta$ as a function of $U$ is plotted in Fig.
\ref{fig_ef6_corr}.
With increasing $U$, the occupancy of the f-states decreases. The double occupancy drops even
faster, resulting in an increase of $\Theta$. In the range of the transition to the $\Gamma$-phase
at $U \approx 1.25 \, t$, the effect of correlations is still negligible, i.e. the transition can also be predicted
using a mean-field decoupling of the interaction Hamiltonian $\hamilton_U$ \cite{interaction-induced_TPT_BHZ-model}.
As the f-orbitals approach half-filling, correlation effects become very pronounced, as is indicated by the measure of correlation $\Theta$
taking values close to 1. Thus we have a regime of well-defined localized moments on the f-orbitals for parameters $U > 8.0 \,t$, and a
small band gap $\Delta_g \ll t$ separating the weakly dis\-per\-sing heavy bands. 
%\xout{This means that the system is clearly in a Kondo insulating state.}
%\comment{The Kondo insulating state is not necessarily linked to the local moment regime.  } 
This becomes manifest also in the spectral function
$A(k,\omega) = -\pi^{-1}\, \op{Im} \op{Tr} G(k,\omega+i \delta)$. In the small frequency limit, we can
expand the self-energy
$\Sigma(\omega) \approx \Sigma(\omega = 0) + \omega \frac{d \Sigma(\omega)}{d \omega} \Bigr |_{\omega = 0}$.
This way we obtain the low-energy spectrum, as shown in Fig. \ref{fig_Aom}a for the $\Gamma$-phase with weak correlations,
and in Fig. \ref{fig_Aom}b with the strongly renormalized, almost flat bands
of the Kondo insulating state, separated by a small gap $\Delta_g \sim T_{coh}$.

%To investigate the extension of the Mott phase in parameter space, we let the local interaction strength take a very large value
% of $U = 10.0 \,t$, and starting from a state inside the $M$-phase at $\eps_f = -1.0\,t$, we decrease $\eps_f$.
%As can be seen in Fig. \ref{fig_Mott_gap}, the band gap size decreases. The system approaches the transition line
%$\Sigma_0 = -\eps_f$ while, at the same time the effective mass starts to increase, as Fig. \ref{fig_Mott_m_eff} shows.
%Right at the transition line, it seems to diverge. The band gap size decreases to zero very rapidly at this point.\\
%In the orbital-selective Mott state the d-band and f-band are completely decoupled. The d-band is in a gap-less, metallic state,
%while the Mott insulating f-band has a Mott gap with a gap size $\sim \mathcal{O}(U)$.
%\begin{center}
%	\includegraphics[width=0.5\textwidth]{plot_Mott_gap_vs_ef.eps}
%	\captionof{figure}{Band gap size on approaching the transition to the Mott phase}
%	\label{fig_Mott_gap}
%\end{center}

%\begin{center}
%	\includegraphics[width=0.5\textwidth]{plot_Mott_m_eff.eps}
%	\captionof{figure}{$\Sigma_0$ and $m_{eff}$ along the transition from the $M$-phase to the Mott phase. The black line represents
%the transition line $\Sigma_0 = -\eps_f$ between $\Gamma$-phase and $M$-phase.}
%	\label{fig_Mott_m_eff}
%\end{center}

\section{Discussion \& Conclusions}
\label{sec_Discussion}
We have described and  studied within the DMFT approximation a model relevant for the understanding of topological Kondo insulators. 
The concept of the topological Hamiltonian \eqref{eq_h_topo} provides a phase diagram,
which serves as a roadmap for the interpretation of the numerical results, as long as
an adiabatic connection to a non-interacting state exists. The low-temperature phase diagram was
found to be very rich. In particular, it is possible to study a transition via an intermediate semi-metallic state
 between two distinct topological states, the $\Gamma$-phase and the $M$-phase, which is driven by the interaction, and which was not observed previously.
% and the simultaneous change in the $Z_2$ topological invariant.
With open boundaries, the two states exhibit zero-energy edge modes crossing at $k_x = 0$ and $k_x = \pm \pi$ respectively. The different nature of
the two topological states was \mbox{evidenced} by establishing the adiabatic connection to the
respective non-interacting states, and by directly calculating the parities corresponding to
the time-reversal invariant momenta. Starting from a trivial insulating state at $U = 0$, we
can equally observe an interaction-driven transition to the $\Gamma$-phase. This aspect is similar
to investigations of the BHZ-Hubbard model. There, the same correlation-driven transition
from a trivial to topological insulator was found \cite{interaction-induced_TPT_BHZ-model,Budich_Topo_Hund_Insulator}.

All observed zero-temperature phases adiabatically connect to non-interacting states. It is however important to note that 
the topological state survives well into the local moment regime.  This regime is characterized by strongly renormalized bands
and  a small band gap  characteristic  of  Kondo insulators.  The  realization of this state of matter in a  simple toy model allows
the  detailed study of the temperature evolution of the single particle spectral function on various topologies  with and without edges.
In particular such studies will provide a detailed understanding  of the emergence of edge states as the temperature 
crosses the relevant energy scales of Kondo physics. This will become of increasing importance once a Kondo insulating state
is realized in Ce-based surface systems or thin films \cite{Reinert_HF_CePt5}.
Our study is not limited to the two dimensional case, and can likewise provide insight into three dimensional topological Kondo insulators.
Again of particular interest is the temperature dependence of the single particle spectral function as well as the nature,
weak or strong, of the topological state. Studies along these lines are presently under progress.

%of this satet 
%All low temperature phases
%In the strong-coupling regime, we see
%signs of a TRS breaking transition, with 
%a dramatic increase of the effective mass, and an almost complete
%suppression of charge fluctuations on the f-levels. Here the f-electrons represent very well
%defined local moments, which are screened by the conduction electrons to form heavy
%quasi-particles. There is a small gap separating the quasi-particle bands, which means that in this
%parameter regime a topological Kondo insulating state is realized.
%The perspectives.  3D weak and strong. Temperature dependence of the single particle spectral functions. 

\begin{acknowledgments}
JW would like to thank the Elite Network of Bavaria for funding.  FFA thanks the DFG for financial support under the grant proposal AS120/6-1 (FOR1162). 
We thank the J\"{u}lich Supercomputing Centre for allocation of CPU time.
\end{acknowledgments}

\bibliography{journal_short,tki}{}
\end{document}